# Evidence of Weyl Fermion Enhanced Thermal Conductivity Under Magnetic Fields in Antiferromagnetic Topological Insulator Mn(Bi$_{1-x}$Sb$_x$)$_2$Te$_4$

Robert A. Robinson
*Department of Physics, Pennsylvania State University, University Park, Pennsylvania, 16802, USA*

Seng Huat Lee[*]
*2D Crystal Consortium, Materials Research Institute,
The Pennsylvania State University, University Park, PA 16802, USA*

Lujin Min
*Department of Materials Science and Engineering,
Pennsylvania State University, University Park, Pennsylvania 16802, USA*

Jinliang Ning and Jianwei Sun
*Department of Physics and Engineering Physics,
Tulane University, New Orleans, Louisiana 70118, USA*

Zhiqiang Mao[†]
*Department of Physics, Pennsylvania State University,
University Park, Pennsylvania, 16802, USA and
2D Crystal Consortium, Materials Research Institute,
The Pennsylvania State University, University Park, PA 16802, USA*

(Dated: April 12, 2023)

We report thermal conductivity and Seebeck effect measurements on Mn(Bi$_{1-x}$Sb$_x$)$_2$Te$_4$ (MBST) with $x = 0.26$ under applied magnetic fields below 50 K. Our data shows clear indications of the electronic structure transition induced by the antiferromagnetic (AFM) to ferromagnetic (FM) transition driven by applied magnetic field as well as significant positive magnetothermal conductivity in the Weyl semimetal state of MBST. Further, by examining the dependence of magnetothermal conductivity on field orientation for MBST and comparison with the magnetothermal conductivity of MnBi$_2$Te$_4$ we see evidence of a contribution to thermal conductivity due to Weyl fermions in the FM phase of MBST. From the temperature dependence of Seebeck coefficient under magnetic fields for MBST, we also observed features consistent with the Fermi surface evolution from a hole pocket in the paramagnetic state to a Fermi surface with coexistence of electron and hole pockets in the FM state. These findings provide further evidence for the field-driven topological phase transition from an AFM topological insulator to a FM Weyl semimetal.

## I. INTRODUCTION

MnBi$_2$Te$_4$ (MBT) has recently garnered a great deal of interest both as the first intrinsic antiferromagnetic (AFM) topological insulator[1–10] and for its ability to host a variety of topological quantum states, such as quantum anomalous Hall insulator[11], axion insulator[12,13], and Chern insulator states[14–16] in 2D thin layers. MBT has also been theoretically predicted to host an ideal time-reversal symmetry breaking type-II Weyl semimetal (WSM) state under applied field in the $H \parallel c$ direction[1,2]. MBT is a Van der Waals material with septuple layers stacked along the crystallographic $c$-axis in a Te-Bi-Te-Mn-Te-Bi-Te configuration[2,11]. The Mn layers have an intralayer ferromagnetic (FM) ordering and stack with alternating magnetic orientation along the $c$-axis to form an interlayer AFM ordering[1–4,8]. MBT has a Néel temperature, $T_N = 25$ K, and undergoes two magnetic transitions under applied magnetic field, at $H_{c1} = 3.57$ T and $H_{c2} = 7.70$ T, at 2 K when $H \parallel c$[4,5,7,8,11,17,18]. The transition at $H_{c1}$ leads to a canted AFM (CAFM) state[4,8,17,19]. The CAFM state transforms into a FM state above $H_{c2}$[4,8,17,19].

The topological properties of MBT originate from the Bi-Te layers; Bi and Te $p_z$ bands invert at the $\Gamma$ point due to spin orbit coupling resulting in an AFM topological insulator state[2,3]. When the AFM phase is polarized to a FM phase by a magnetic field parallel to the $c$-axis, the topological insulator state is predicted to evolve into a an ideal type-II Weyl state with strongly tilted Weyl cones[2]. Recent theoretical studies further predict such a Weyl state can be tuned by the field orientation, but disappears as the field is rotated to the in-plane direction[20]. However, pristine MBT does not exhibit WSM behavior in the FM phase driven by the $c$-axis magnetic field because the Weyl nodes are too far from the Fermi surface[17].

Recent work[17,21–23] has shown that it is possible to tune the chemical potential of MBT via doping with Sb on the Bi site in order to bring the Fermi level to the Weyl nodes[17]. While MBT is electron doped[8,17,19], as the Sb concentration increases, MBST's chemical poten-tial is tuned from the bulk conduction band to the bulk valence band passing through the charge neutral point near $x = 0.26$[17]. Our prior work has shown that the predicted ideal WSM state is accessible in the lightly hole doped samples with $x = 0.26$. This is revealed through Hall resistivity, anomalous Hall effect, and $c$-axis magnetoresistivity measurements. These measurements demonstrate that the AFM-to-FM transition induces an electronic structure transition and unveils typical transport signatures of a WSM, including a large intrinsic anomalous Hall effect and chiral anomaly[17]. The Weyl state in MBST is of particular interest because it is the least complicated possible manifestation of a Weyl phase, hosting only one pair of Weyl nodes at the Fermi level and having no interference from other trivial bands near the Fermi level[2,17]. Such an ideal Weyl state has long been sought in a condensed matter system, since its simplicity makes it valuable for further understanding Weyl fermion physics. However, it has not been observed in any other materials prior to our demonstration of its existence in MBST.

In this work we have measured the thermal conductivity and Seebeck coefficient of lightly hole doped MBST, $x = 0.26$, with $H \parallel c$ and $H \perp c$ as well as MBT with $H \parallel c$ at various temperatures below 50 K. In doing so we have observed further evidence of the electronic transition induced by the field driven AFM-to-FM transition. Furthermore, we have observed a substantial enhancement of thermal conductivity in lightly hole doped MBST above $H_{c2}$ that we accredit to a contribution due to Weyl fermions, indicating that these exotic particles can play an important role in heat conduction in a material. We also observed a more than linear suppression of the Seebeck coefficient with decreasing temperature in response to the paramagnetic (PM)-to-FM crossover-like transitions under high magnetic fields. This behavior is consistent with a transition from a hole Fermi pocket in the PM phase to a combined hole and electron pocket state in the FM phase and further supports the presence of the WSM state in MBST $x = 0.26$.

## II. EXPERIMENTAL DETAILS

MnBi$_2$Te$_4$ and Mn(Bi$_{1-x}$Sb$_x$)$_2$Te$_4$ were synthesized using the methods previously established[4,17]. Phase purity of the samples was checked via X-ray diffraction and Sb content of each sample was determined by energy-dispersive X-ray spectroscopy. No samples used showed any indication of impurity.

Thermal and thermoelectric measurements were carried out via the 4 wire method (see Fig. 1a inset) using the thermal transport option in a Physical Property Measurement System (PPMS, Quantum Design). Thermal conductivity was measured by heating one end of the sample with cross-sectional area, $A$, using a resistive heater with a power, $W$, while fixing the temperature at the other end via a cold foot. This configuration creates a time dependent temperature gradient which the software then uses to compute a steady state temperature difference, $\Delta T$. This process is measured by 2 thermometer probes separated by a known distance, $\Delta d$. From these values, thermal conductivity, $\kappa$, can be determined by the equation below[24]:

$$\kappa = \frac{W \Delta d}{A \Delta T} \left[\frac{W}{m\ K}\right] \quad (1)$$

The Seebeck coefficient is measured by using the same method as thermal conductivity to compute $\Delta T$, and measuring potential difference, $\Delta V$, using a volt meter. The Seebeck coefficient is then computed by[25]:

$$S = \frac{\Delta V}{\Delta T} \left[\frac{\mu V}{K}\right] \quad (2)$$

Thermal conductivity and Seebeck coefficient measurements of MBST were done on different samples with nearly identical chemical composition ($x = 0.26$). Differences in sample geometry can affect the quality of the measured data, with thicker samples favoring thermal conductivity measurements by allowing for a higher wattage to be applied to the sample. As such, a thicker sample was used for thermal conductivity measurements. However, Seebeck effect measurements require a greater temperature gradient; thus a thinner sample was used. Due to the fragility of the thin sample used to measure the Seebeck effect, a Teflon substrate was used to make the sample more robust. The Teflon was attached to the sample using double sided tape and did not make contact with the sample leads. Teflon is used as it is both a good thermal and electrical insulator; tests on reference samples have shown that it does not meaningfully impact the data (see Fig. S1).

In-plane resistivity, $\rho_{xx}$, measurements were conducted via the standard 4-probe method using the resistivity option in a PPMS. $\rho_{xx}$ measurements used the same sample as the thermal conductivity measurements. Magnetic susceptibility measurements were conducted via a Quantum Design Magnetic Property Measurement System using the same sample as the Seebeck coefficient measurements.

First-principles calculations based on density functional theory[26] are performed using the Vienna Ab-initio Simulation Package (VASP)[27] with the projector-augmented wave (PAW) method[28,29]. The strongly-constrained and appropriately-normed (SCAN) meta-GGA developed in 2015[30,31] has shown superior performance in description of different chemical bonds and transition metal compounds[30–35]. In this work, we used a recently modified version of SCAN (r2SCAN[35]) with improved performances especially in numerical stability[36–38]. The state-of-art D4 dispersion correction method[39,40] was combined with r2SCAN for a better description of van der Waals interactions. The PAW method is employed to treat the core ion-electron interaction and the valence configurations are taken as Mn: $3p^64s^13d^6$, Bi: $6s^26p^3$, Te: $5s^25p^4$ and Sb: $5s^25p^3$. An energy cutoff of 520 eV is used to truncate the plane wave basis, together with a high real space grid setting (PREC = high; ENCUT = 520; ENAUG = 2000). We use $\Gamma$-centered meshes with a spacing threshold of KSPACING = 0.15 Å$^{-1}$ for K-space sampling. Geometries of MnBi$_2$Te$_4$ and Mn(Bi$_{0.75}$Sb$_{0.25}$)$_2$Te$_4$ were allowed to relax without considering spin-orbit coupling (SOC) until the maximum ionic forces were below a threshold of 0.001 eV Å$^{-1}$.

## III. RESULTS & DISCUSSION

### A. Thermal Conductivity

Thermal conductivity measurements were conducted on Mn(Bi$_{1-x}$Sb$_x$)$_2$Te$_4$, $x = 0.26$, from 0-9 T with $H \parallel c$ at selected temperatures above and below $T_N$. This data is plotted as magnetothermal conductivity,

$$\frac{\Delta\kappa}{\kappa(0T)} = \frac{\kappa(B) - \kappa(0T)}{\kappa(0T)} \qquad (3)$$

in Fig. 1a. Below $T_N$ there are two clear transitions that correspond to the magnetic transitions in MBST. These transitions are marked with arrows on the plot and tend towards lower field with increasing temperature. In order to verify that the observed behavior is a result of the magnetic transitions, we used this data to construct a phase diagram and compared it with one constructed using magnetoresistivty data in ref.[17] (Fig. 1b). Comparing these results, the phase diagrams are consistent and both $H_{c1}$ and $H_{c2}$ trend towards 0 as they approach $T_N$. At low temperatures both data sets approach the reported values for $H_{c1}$ and $H_{c2}$[4,17]. The slightly lower critical field values for the thermal conductivity data are likely due to the fact that different samples were used for each measurement and while they had very close chemical compositions, one may have been slightly more Sb doped than the other. The agreement between these results indicates the thermal conductivity of MBST is sensitive to its spin-flop transition. Therefore, an in depth inspection of the data could lead to deep insights about the physics involved.

Below $T_N$ there are 3 distinct regimes punctuated by $H_{c1}$ and $H_{c2}$ and each exhibits a different behavior. Below $H_{c1}$, the thermal conductivity decreases sharply with increasing field similar to MBT[18,41]. This decrease is associated with an increase in phonon-magnon scattering caused by field-driven increase in the overlap of the phonon and magnon energy bands in momentum space[41]. The suppression could also be due, in part, to suppression of the magnon contribution to thermal conductivity, $\kappa_{mag}$, with applied field[18]. In between $H_{c1}$ and $H_{c2}$ the thermal conductivity is close to constant with a slight increase as field increases. There are two proposed explanations for this behavior in MBT, one is that the shared phonon-magnon phase space is field independent in the CAFM state so the phonon-magnon scattering is unaffected by the changing field[41]. The other explanation is that in this region there is still suppression of $\kappa_{mag}$ but there is also an enhancement in the phonon contribution to thermal conductivity, $\kappa_{ph}$, due to decreased phonon-magnon scattering[18]. Above $H_{c2}$, the thermal conductivity increases drastically displaying positive magnetothermal conductivity, increasing by about 10% in between $H_{c2}$ and 9 T. Since the in-plane magnetoresistivity of the lightly hole-doped MBST sample with $x = 0.26$ exhibits positive magnetoresistivity as presented below (Fig. 3c), the observed positive magnetothermal conductivity can not be understood in terms of Wiedemann-Franz law.

MBT also experiences an increase in thermal conductivity above $H_{c2}$, however $\kappa(B)$ never exceeds the 0 field value. In MBT the increase is attributed to a widening of the gap between the phonon and magnon bands in k-space, reducing the phonon-magnon scattering[41]; an overall suppression of the number of magnons above $H_{c2}$ could also reduce the phonon scattering[18]. While the same effect does likely play a role in MBST, it is unlikely to be the explanation for the positive overall magnetothermal conductivity. The magnetic and crystal structure of MBST is very similar to MBT, and, as will be discussed shortly, phonons appear to play less of a role in the total thermal conductivity of MBST as compared to MBT. Therefore it would not make sense for the reduced phonon magnon scattering to lead to a greater enhancement in MBST. The one significant difference between the FM states of MBT and MBST is the presence of the Weyl state in MBST implying that there could be a contribution to $\kappa$ associated with the WSM state present in MBST. We will address this in more detail below.

Above $T_N$ there are no transitions so the thermal conductivity slightly increases at 35 K and slightly decreases at 50 K with increasing field (Fig. 1a bottom panel). The slight increase is likely due to suppression of phonon scattering by the magnons associated with the short-range, intrapalanar FM ordering[18,41] that exists when the material is just above $T_N$ while the slight decrease at 50 K is likely due to a slight suppression of the charge carrier contribution to thermal conductivity, $\kappa_e$, caused by the Lorentz force[42,43].

Along with the thermal conductivity measurements conducted on Mn(Bi$_{1-x}$Sb$_x$)$_2$Te$_4$ $x = 0.26$ from 0-9 T with $H \parallel c$, measurements on the same sample with $H \perp c$ and on MnBi$_2$Te$_4$ with $H \parallel c$ were also performed (Fig. 2). These data allow for a comparison of the thermal conductivities as a function of magnetic field. From these comparisons, we can isolate behavior in the magnetothermal conductivity associated with the WSM state.

The WSM state in MBST exists above $H_{c2}$ when the component of the magnetic field parallel to the crystallographic $c$ direction is nonzero, if the field is entirely in the $ab$ plane then the WSM state does not occur[17,20]. By measuring the thermal conductivity of the same sample with $H \parallel c$ and $H \perp c$, we can separate out behavior associated with the WSM state from other intrinsic behavior of the material. At 5 K (Fig. 2a) the difference is striking; the measurements with $H \parallel c$ have distinct phase transitions and positive magnetothermal conductivity above 8 T, while those with $H \perp c$ exhibit no inflection points and negative magnetothermal conductivity. While both magnetic field orientations lead to a transition from AFM to FM states, in the $H \perp c$ configuration this transition happens more smoothly and $H_{c2} \approx 10$ T, so the lack of clear inflection points is not a surprise and is consistent with previously reported results[17]. In order to compare the thermal conductivity in the FM phases driven by the in-plane and out-of-plane magnetic fields, we increased the temperature to 10 K (Fig. 2b) which decreases the $H_{c2}$ of the $H \perp c$ to around 8 T. At 10 K we now see an enhancement in the FM phase for both measurements. However, the enhancement in $H \parallel c$ is still substantially larger than that of $H \perp c$. Both the $H \parallel c$ and $H \perp c$ configurations are in the FM phase above $H_{c2}$ at 10 K, the only difference is the presence of a WSM state in the $H \parallel c$ configuration. This implies that the WSM state could be responsible for the larger enhancement in magnetothermal conductivity, implying that the Weyl fermions have greater electronic contributions to thermal conductivity than normal electrons. We find further evidence for this scenario by comparing the thermal conductivity data of MBST with those of MBT. From the MBST data sets alone we cannot rule out suppressed phonon scattering as the cause of the thermal conductivity enhancement above $H_{c2}$. Figure 2c shows thermal conductivity above $T_N$ where the direction of the magnetic field should not have an impact on the thermal conductivity of the PM state, and the MBST data sets are indeed nearly identical as we would expect.

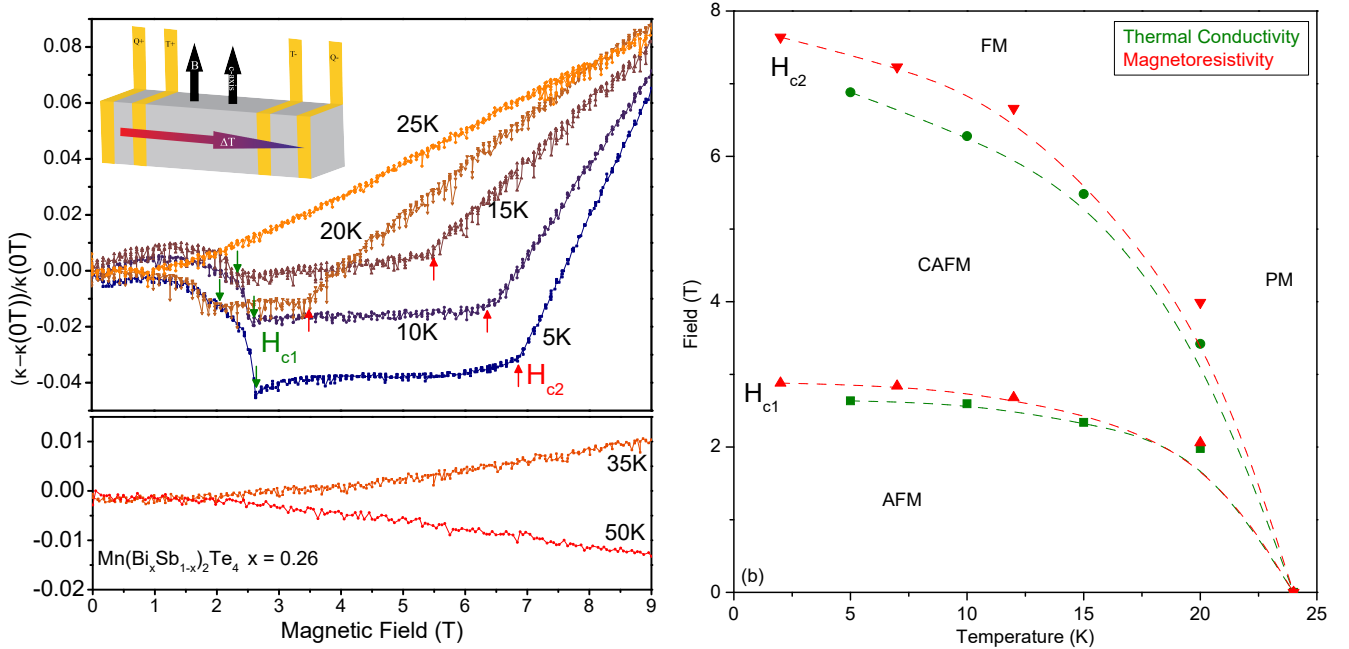

FIG. 1: **(a)** Normalized thermal conductivity data of $Mn(Bi_{1-x}Sb_x)_2Te_4$ $x = 0.26$. Data below $T_N$ is plotted separately from data above $T_N$, $H_{c1}$ for each temperature is indicated by a green, downward pointing arrow, $H_{c2}$ is indicated by an red, upward pointing arrow. There is substantial enhancement of the thermal conductivity above $H_{c2}$. **Inset:** Schematic diagram of a thermal conductivity measurement. The gray bar represents the sample with 4 copper leads attached. From left to right the leads are heater source, hot zone thermometer, cold zone thermometer, and cold sink. For these measurements $H_{applied} \parallel c-axis$. **(b)** Phase diagram of $Mn(Bi_{1-x}Sb_x)_2Te_4$ $x = 0.26$ below $T_N$. Two independent phase diagrams were constructed, one using thermal conductivity data (green) and on using magnetoresistivity data (red)[17]. The results for $H_{c1}$ and $H_{c2}$ are in good agreement between the data sets. The small discrepancy is likely due to the measurements being done on different samples with very similar doping levels.

MBT and MBST at $x = 0.26$ have similar crystal structures, magnetic structures, and magnetic transitions. The only notable differences are the presence of the WSM state in MBST and slightly more disordered lattice in MBST due to the Sb doping. By comparing the behavior of MBT and MBST under the same conditions we can determine what effect these differences have on thermal conductivity. Above $T_N$ (Fig. 2c) we see a greater enhancement in MBT than we do in MBST. In MBT this enhancement is associated with a decrease in phonon-magnon scattering as noted above[18,41]; it is likely that the same is true of MBST. While both samples see an improvement in thermal conductivity under magnetic field, the larger increase in MBT indicates $\kappa_{ph}$ is a larger overall proportion of the thermal conductivity in MBT, meaning that reducing phonon scattering has more of an impact. This makes sense as MBST's more disordered lattice would increase phonon scattering, decreasing the relative contribution of $\kappa_{ph}$ and the FM phase of MBST has much higher carrier mobility due to the presence of the Weyl state[17] which would enhance the charge carrier contribution to thermal conductivity.

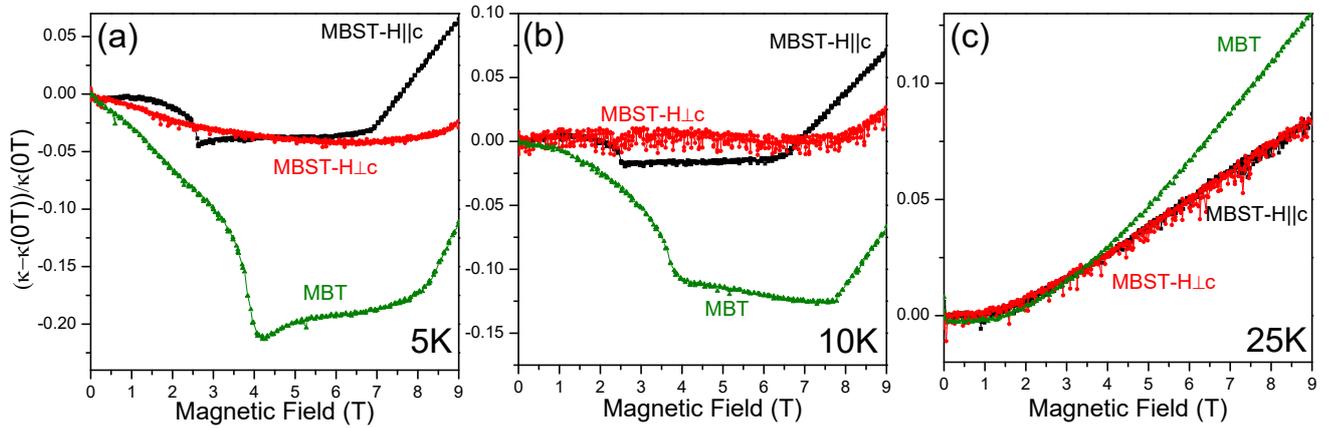

FIG. 2: Comparison of normalized thermal conductivity of $Mn(Bi_{1-x}Sb_x)_2Te_4$ (MBST) $x = 0.26$ with $H \parallel c$ and with $H \parallel ab-plane$, and $MnBi_2Te_4$ (MBT) with $H \parallel c$ at (a) 5 K (b) 10 K, and (c) 25 K.

At 5 K (Fig. 2a) MBT and MBST have very similar behavior below $H_{c2}$, dropping precipitously as $H \to H_{c1}$ then levelling off between $H_{c1}$ and $H_{c2}$. The drop below $H_{c1}$ is attributed to an increase in the intersection of the phonon and magnon bands in k-space[41], so the larger drop in MBT suggests that phonons play a larger role in the total thermal conductivity, agreeing with our 25 K results. It is also possible that the magnons play less of a role in the thermal conductivity of MBST due to an increase in lattice disorder caused by the Sb doping.

Above $H_{c2}$ there is a difference, while the thermal conductivity of both samples increases linearly, MBT has a negative overall magnetothermal conductivity while MBST has a positive magnetothermal conductivity. In MBT this increase is associated with phonon-magnon scattering suppression[18,41]; it is likely that this also contributes to the behavior in MBST. However, in MBT the enhancement due to reduced scattering is of the same magnitude as the decrease in thermal conductivity below $H_{c1}$ because both are related to a change in the overlap between the phonon and magnon bands. So, in MBST we would expect any enhancement from phonon-magnon scattering suppression to similarly be of the same magnitude as the suppression below $H_{c1}$ because once the phonon and magnon bands no longer overlap at high field, this effect can not further enhance the thermal conductivity. Given that we observe a substantially larger enhancement, it follows that some other effect must be involved.

The bipolar effect, in which coupled electrons and holes form, travel through the lattice from the hot to cold end, then annihilate, has been shown to enhance thermal conductivity in a way similar to our results[44–46], and given that lightly hole doped MBST, unlike MBT, hosts both electrons and holes in the FM state, it is possible that this is the source of the extra contribution. However, we find this explanation unlikely to be correct for two reasons, the temperature dependence of the observed enhancement and the relative sizes of the electron and hole pockets in MBST. In general, the bipolar effect is expected to manifest at high temperatures[44,45], only manifesting in MBT above 115 K[46] and is expected to increase with temperature[44]. However, the enhancement in MBST appears at low temperatures and increases with decreasing temperature. Further, the electron pockets in MBST are substantially smaller than the hole pockets[2,17,22,47] and the bipolar effect depends on coupled electrons and holes moving through the lattice[44,45] and so would be very limited by the size of the electron pockets in MBST.

Next, we consider the effect caused by the possible evolution of magnetic fluctuations from MBT to MBST. If we assume that Sb substitution for Bi in MBST leads to larger magnetic fluctuations than in MBT, the enhancement of magnetothermal conductivity due to increasing the applied field would be greater. Weaker single ion anisotropy (SIA) or interlayer coupling (IC) could cause MBST to have larger magnetic fluctuations. To check for a difference in SIA between MBT and MBST $x = 0.26$, we measured the magnetization of both samples under applied magnetic fields with $H \parallel c$ ($M_c$) and $H \parallel ab$ ($M_{ab}$). We then plotted the ratio of the magnetizations, $\frac{M_c}{M_{ab}}$, (Fig. S2) as greater values of this ratio would indicate larger SIA. We compared these values for MBT and MBST $x = 0.26$ and found that SIA was slightly larger in the CAFM and FM phases of the doped sample indicating that this could not explain the enhanced magnetothermal conductivity. To check the (IC) strength of MBT and MBST $x = 0.26$ we computed it for both materials and found that the A-type AFM phase is calculated to be 1.94 (2.75) meV/f.u lower in energy than the FM phase for the pristine MBT, without (with) SOC considered, and 2.08 (2.68) meV/f.u. for MBST $x = 0.25$. So, the interlayer Mn-Mn coupling strength will be 0.32 (0.46) meV for pristine MBT without (with) SOC considered, and 0.35 (0.45) meV for MBST $x = 0.25$. These results indicate that the IC strength is not substantially different between MBT and MBST. This is further evidenced by the fact that the $c$ lattice parameter is nearly the same between these two compounds[21].

It is also worth considering the possibility that reduced charge carrier concentration in MBST could reduce the amount of phonon-electron scattering, magnifying the effect of changes in phonon-magnon scattering. MBST with $x = 0.26$ has a much lower charge carrier density than MBT in the paramagnetic state. However, MBST undergoes an electronic structure transition above $H_{c2}$[17] that does not occur in MBT, and above this transition MBST hosts both electron and hole pockets making the carrier density difficult to evaluate from two-band model fitting. Therefore, it is not safe to assume that the carrier concentration of MBST in the FM state is much lower than in MBT. Further, in the reported magnetothermal conductivity data we normalized our results to the 0 field value, so if there was any overall shift in magnitude, that has been accounted for by the normalization. Furthermore, if we were to assume that decreased phonon-magnon scattering should have a larger effect on the magnetothermal conductivity of MBST than MBT, then we should expect both a more extreme suppression and enhancement of magnetothermal conductivity in the AFM and FM phases of MBST respectively, because both are due to changes in phonon-magnon scattering. However, we see a smaller suppression in the AFM phase but a larger enhancement in the FM phase (Fig. 2a & b), which is not consistent with this assumption. Prior work has demonstrated that the suppression of thermal conductivity in MBT is due to the magnon-phonon scattering, rather than the change of electron-phonon scattering[18,41]; prior magnetotransport studies have shown electron-magnon scattering is suppressed above $H_{c1}$ in MBT, while in MBST electron-magnon scattering is overwhelmed by the chiral anomaly effect of the Weyl state[8,17]. Therefore, it is most reasonable to attribute the positive magnetoconductivity above $H_{c2}$ in MBST to the Weyl Fermions' contributions.

Having excluded all other reasonable explanations, we conclude that we are observing a contribution to the thermal conductivity due to Weyl fermions. Given that the overall enhancement above $H_{c2}$ is linear, and we know any contribution from reduced phonon-magnon scattering would be linear, this implies that any enhancement from Weyl fermions must also contribute a linear term. As for why the Weyl contribution increases linearly with field, we believe that as the magnetic field is increased more electrons become available at the Fermi level allowing for increased thermal conduction. An increase in available Weyl fermions with increasing field is consistent with established physics. Consider how the Landau levels (LL) behave under applied field: the Fermi level of MBST $x = 0.26$ is near the Weyl nodes, meaning that the $0^{th}$ LL is pinned to the node and cannot move[48]. However, as the field is increased the higher LLs move further from the Fermi level increasing the degeneracy of the $0^{th}$ LL[48], making more charge carriers available to act as Weyl fermions in the WSM state.

### B. Seebeck Coefficient

Seebeck coefficient data was collected for Mn(Bi$_{1-x}$Sb$_x$)$_2$Te$_4$ $x = 0.26$ for fields from 0-9 T in between $5 - 50$ K (Fig. 3a). To see how the Seebeck coefficient responds to the magnetic transitions, we also measured temperature dependence of resistivity (Fig. 3c) and magnetic susceptibility(Fig. 3d) at various magnetic fields, from which we extracted the field dependence of the magnetic transition temperature $T_N$ (Fig. 3b). While the PM-to-AFM (CAFM) transitions at lower fields ($< 7$ T) can be clearly resolved as denoted by the arrows in Fig. 3c&d, the PM-to-FM transition at 7 T or 9 T is a crossover-like broad transition. We have added the color map of magnetic transitions to Fig. 3a to better see how the Seebeck coefficient is coupled to the magnetic transitions. The Seebeck data below 5 T indicate that MBST is hole dominated but close to the charge neutral point[17], leading to a small, positive Seebeck coefficient that decreases linearly with decreasing temperature. Unlike in MBT[18,41], we do not observe a feature in the Seebeck coefficient at $T_N$; this is due to MBST, $x = 0.26$, being near the charge neutral point. The Seebeck effect depends on the charge carrier density[25], so in a material with very few charge carriers, like our sample, the Seebeck coefficient is only measurably effected by major changes in the electronic structure. Hence, effects that only have a small impact on the electronic state, like the magnetic transition at $T_N$, are washed out. At fields above 5 T the Seebeck coefficient decreases linearly until the material enters the Weyl state at which point an electron pocket opens up at the Fermi surface[17]. In conjunction with the hole pocket, this electron pocket's presence further suppresses the Seebeck coefficient. This leads to a greater than linear decrease in the Seebeck coefficient once the material enters the FM Weyl state. Fit lines are included in Fig. 3a; data at 5 T where the system enters the CAFM state below $T_N$ also shows clear deviation below $T_N$. This is because the WSM state starts to appear in the CAFM state as discussed in our prior work[17]. Therefore, the Seebeck coefficient data provides additional support of the coexistence of electron and hole pockets in the FM state. The results of these Seebeck coefficient measurements agree with our thermal conductivity results demonstrating thermal and thermoelectric response to the field driven WSM state in MBST. Additionally, from Fig. 3b, it can be seen that the in-plane resistivity ($\rho_{xx}$) of the MBST sample ($x = 0.26$) measured under magnetic fields along the c-axis increases with magnetic field (i.e. positive magnetoresistivity) and its temperature dependence displays metallic-like behavior in the AFM phase at zero or low fields but insulating like behavior in the CAFM or FM phase at higher fields. These features also provide additional support for the argument of the transition from the AFM topological insulator to the FM Weyl semimetal state, as discussed in our earlier work[17].

## IV. CONCLUSION

By measuring the thermal conductivity of $Mn(Bi_{1-x}Sb_x)_2Te_4$ $x = 0.26$ from 0-9 T with $H \parallel c$ we have been able to probe the magnetic phase transitions in MBST as well as observe a large enhancement in the thermal conductivity above $H_{c2}$. By comparing these results with measurements of the same sample with $H \perp c$ and of $MnBi_2Te_4$ with $H \parallel c$ we have presented strong evidence that Weyl fermions play a role in heat conduction and contribute meaningfully to the thermal conductivity of MBST. Through measurements of the Seebeck coefficient we have found further experimental support of the field-driven Weyl semimetal state.

### Acknowledgments

This work is primarily supported by the National Science Foundation through the Penn State 2D Crystal Consortium-Materials Innovation Platform (2DCC-MIP) under NSF Cooperative Agreement DMR-2039351. Z.Q.M. also acknowledges the support from the US National Science Foundation under grant DMR 2211327. J.W.S. acknowledges the support from the US Department of Energy under grant DE-SC0014208 for the computational work.

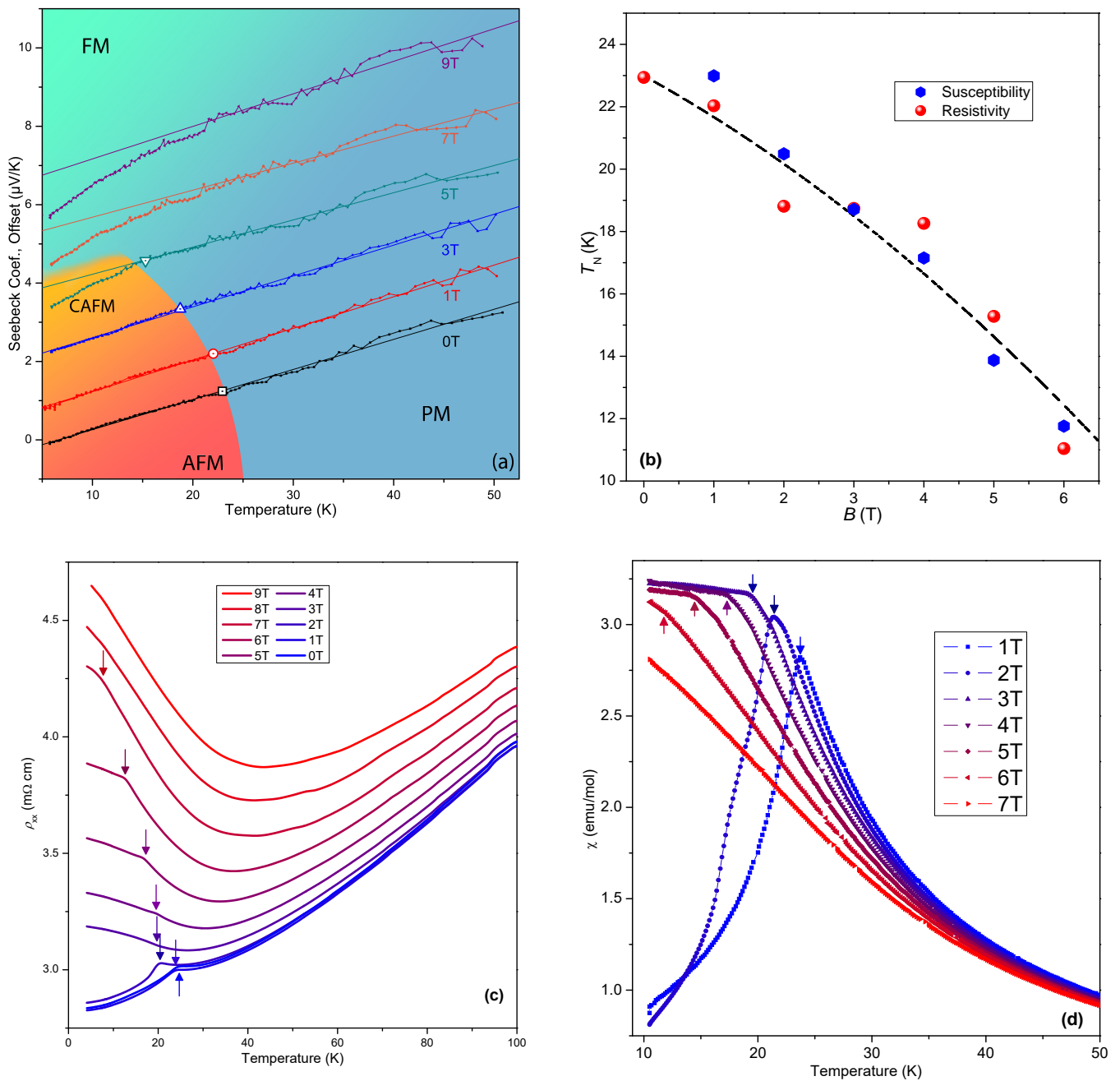

FIG. 3: **(a)** Seebeck coefficient data for Mn(Bi$_{1-x}$Sb$_x$)$_2$Te$_4$ $x = 0.26$ with $H \parallel c$ at selected field values with associated linear fits. 0T data is at appropriate scale, data under applied fields is offset for clarity. Above $T_N$, the sample exhibits linear behavior at all field values. Below $T_N$, the data above 3 T shows a clear deviation from linearity associated with the electronic transition from a hole Fermi pocket in the PM state to the coexistence of electron and hole pockets in the CAFM/FM state. Large symbols are used to indicate magnetic transition temperature derived from resistivity and magnetic susceptibility. Background is colored to show different magnetic states as a function of temperature and field. **(b)** Magnetic transition temperature as a function of applied field derived from resistivity and magnetic susceptibility. **(c)** $\rho_{xx}$ vs. T data at different applied field strengths with $H \parallel c$, arrows indicate transition temperatures, $T_N$, computed from the first derivative. **(d)** $\chi$ vs. T data at different applied field strengths with $H \parallel c$, arrows indicate transition temperatures, $T_N$, computed from the first derivative.
8

* Email: shl12@psu.edu
† Email: zim1@psu.edu

[1] D. Zhang, M. Shi, T. Zhu, D. Xing, H. Zhang, and J. Wang, Phys. Rev. Lett. **122**, 206401 (2019).
[2] J. Li, Y. Li, S. Du, Z. Wang, B.-L. Gu, S.-C. Zhang, K. He, W. Duan, and Y. Xu, Science Advances **5**, eaaw5685 (2019).
[3] M. M. Otrokov, I. I. Klimovskikh, H. Bentmann, D. Estyunin, A. Zeugner, Z. S. Aliev, S. Gaß, A. Wolter, A. Koroleva, A. M. Shikin, et al., Nature **576**, 416 (2019).
[4] J.-Q. Yan, Q. Zhang, T. Heitmann, Z. Huang, K. Chen, J.-G. Cheng, W. Wu, D. Vaknin, B. C. Sales, and R. J. McQueeney, Physical Review Materials **3**, 064202 (2019).
[5] P. M. Sass, W. Ge, J. Yan, D. Obeysekera, J. Yang, and W. Wu, Nano Letters **20**, 2609 (2020).
[6] R. Vidal, H. Bentmann, T. Peixoto, A. Zeugner, S. Moser, C.-H. Min, S. Schatz, K. Kißner, M. Ünzelmann, C. Fornari, et al., Physical Review B **100**, 121104 (2019).
[7] Y. Chen, L. Xu, J. Li, Y. Li, H. Wang, C. Zhang, H. Li, Y. Wu, A. Liang, C. Chen, et al., Physical Review X **9**, 041040 (2019).
[8] S. H. Lee, Y. Zhu, Y. Wang, L. Miao, T. Pillsbury, H. Yi, S. Kempinger, J. Hu, C. A. Heikes, P. Quarterman, et al., Physical Review Research **1**, 012011 (2019).
[9] Y.-J. Hao, P. Liu, Y. Feng, X.-M. Ma, E. F. Schwier, M. Arita, S. Kumar, C. Hu, M. Zeng, Y. Wang, et al., Physical Review X **9**, 041038 (2019).
[10] H. Li, S.-Y. Gao, S.-F. Duan, Y.-F. Xu, K.-J. Zhu, S.-J. Tian, J.-C. Gao, W.-H. Fan, Z.-C. Rao, J.-R. Huang, et al., Physical Review X **9**, 041039 (2019).
[11] Y. Deng, Y. Yu, M. Z. Shi, Z. Guo, Z. Xu, J. Wang, X. H. Chen, and Y. Zhang, Science **367**, 895 (2020).
[12] C. Liu, Y. Wang, H. Li, Y. Wu, Y. Li, J. Li, K. He, Y. Xu, J. Zhang, and Y. Wang, Nature materials **19**, 522 (2020).
[13] A. Gao, Y.-F. Liu, C. Hu, J.-X. Qiu, C. Tzschaschel, B. Ghosh, S.-C. Ho, D. Bérubé, R. Chen, H. Sun, et al., Nature **595**, 521 (2021).
[14] J. Cai, D. Ovchinnikov, Z. Fei, M. He, T. Song, Z. Lin, C. Wang, D. Cobden, J.-H. Chu, Y.-T. Cui, et al., Nature communications **13**, 1 (2022).
[15] J. Ge, Y. Liu, J. Li, H. Li, T. Luo, Y. Wu, Y. Xu, and J. Wang, National science review **7**, 1280 (2020).
[16] C. Liu, Y. Wang, M. Yang, J. Mao, H. Li, Y. Li, J. Li, H. Zhu, J. Wang, L. Li, et al., Nature Communications **12**, 1 (2021).
[17] S. H. Lee, D. Graf, L. Min, Y. Zhu, H. Yi, S. Ciocys, Y. Wang, E. S. Choi, R. Basnet, A. Fereidouni, et al., Physical Review X **11**, 031032 (2021).
[18] H. Zhang, C. Xu, S. Lee, Z. Mao, and X. Ke, Physical Review B **105**, 184411 (2022).
[19] S. H. Lee, D. Graf, R. A. Robinson, and Z. Mao (2022).
[20] P. Wang, J. Ge, J. Li, Y. Liu, Y. Xu, and J. Wang, The Innovation **2**, 100098 (2021).
[21] J.-Q. Yan, S. Okamoto, M. A. McGuire, A. F. May, R. J. McQueeney, and B. C. Sales, Physical Review B **100**, 104409 (2019).
[22] Q. Jiang, C. Wang, P. Malinowski, Z. Liu, Y. Shi, Z. Lin, Z. Fei, T. Song, D. Graf, S. Chikara, et al., Physical Review B **103**, 205111 (2021).
[23] B. Chen, F. Fei, D. Zhang, B. Zhang, W. Liu, S. Zhang, P. Wang, B. Wei, Y. Zhang, Z. Zuo, et al., Nature communications **10**, 1 (2019).
[24] F. P. Incropera, D. P. DeWitt, T. L. Bergman, A. S. Lavine, et al., *Fundamentals of heat and mass transfer*, vol. 6 (Wiley New York, 1996).
[25] S. J. Blundell and K. M. Blundell, *Concepts in thermal physics* (Oxford University Press on Demand, 2010).
[26] W. Kohn and L. J. Sham, Physical review **140**, A1133 (1965).
[27] G. Kresse and J. Furthmüller, Physical review B **54**, 11169 (1996).
[28] P. E. Blöchl, Physical review B **50**, 17953 (1994).
[29] G. Kresse and D. Joubert, Physical review b **59**, 1758 (1999).
[30] J. Sun, R. C. Remsing, Y. Zhang, Z. Sun, A. Ruzsinszky, H. Peng, Z. Yang, A. Paul, U. Waghmare, X. Wu, et al., Nature chemistry **8**, 831 (2016).
[31] J. Sun, A. Ruzsinszky, and J. P. Perdew, Physical review letters **115**, 036402 (2015).
[32] J. Sun, B. Xiao, Y. Fang, R. Haunschild, P. Hao, A. Ruzsinszky, G. I. Csonka, G. E. Scuseria, and J. P. Perdew, Physical review letters **111**, 106401 (2013).
[33] D. A. Kitchaev, H. Peng, Y. Liu, J. Sun, J. P. Perdew, and G. Ceder, Physical Review B **93**, 045132 (2016).
[34] H. Peng and J. P. Perdew, Physical Review B **96**, 100101 (2017).
[35] J. W. Furness, Y. Zhang, C. Lane, I. G. Buda, B. Barbiellini, R. S. Markiewicz, A. Bansil, and J. Sun, Communications Physics **1**, 1 (2018).
[36] J. W. Furness, A. D. Kaplan, J. Ning, J. P. Perdew, and J. Sun, The journal of physical chemistry letters **11**, 8208 (2020).
[37] J. Ning, J. W. Furness, and J. Sun, Chemistry of Materials **34**, 2562 (2022).
[38] J. Ning, M. Kothakonda, J. W. Furness, A. D. Kaplan, S. Ehlert, J. G. Brandenburg, J. P. Perdew, and J. Sun, Physical Review B **106**, 075422 (2022).
[39] E. Caldeweyher, S. Ehlert, A. Hansen, H. Neugebauer, S. Spicher, C. Bannwarth, and S. Grimme, The Journal of chemical physics **150**, 154122 (2019).
[40] S. Ehlert, U. Huniar, J. Ning, J. W. Furness, J. Sun, A. D. Kaplan, J. P. Perdew, and J. G. Brandenburg, The Journal of Chemical Physics **154**, 061101 (2021).
[41] D. Vu, R. Nelson, B. L. Wooten, J. Baker, J. E. Goldberger, and J. P. Heremans, arXiv preprint arXiv:2203.08032 (2022).
[42] S. N. Guin, K. Manna, J. Noky, S. J. Watzman, C. Fu, N. Kumar, W. Schnelle, C. Shekhar, Y. Sun, J. Gooth, et al., NPG Asia Materials **11**, 1 (2019).
[43] R. A. Robinson, L. Min, S. H. Lee, P. Li, Y. Wang, J. Li, and Z. Mao, Journal of Physics D: Applied Physics **54**, 454001 (2021).
[44] R. Kumar and R. Singh, *Thermoelectricity and Advanced Thermoelectric Materials* (Woodhead Publishing, 2021).
[45] H. Wang, X. Luo, K. Peng, Z. Sun, M. Shi, D. Ma, N. Wang, T. Wu, J. Ying, Z. Wang, et al., Advanced Functional Materials **29**, 1902437 (2019).
[46] H. Wang, X. Luo, M. Shi, K. Peng, B. Lei, J. Cui, D. Ma, W. Zhuo, J. Ying, Z. Wang, et al., Physical Review B **103**, 085126 (2021).
[47] Y. Wang, arXiv preprint arXiv:2103.12730 (2021).
[48] J. Hu, S.-Y. Xu, N. Ni, and Z. Mao, Annual Review of

Materials Research **49**, 207 (2019).